\documentstyle[twocolumn,aps,epsf]{revtex}

\def\i{{\rm i}}
\def\d{{\rm d}}
\def\e{{\rm e}}
\def\vector#1{{\bf #1}}

\def\vk{{\vector k}}
\def\vq{{\vector q}}

\def\dps{\displaystyle}
\def\vF{{v_{\rm F}}}

\def\Tc{{T_{\rm c}}}

\def\qbar{{\bar q}}

\def\kB{{k_{\rm B}}}

\def\BETSFeCl{\mbox{${\rm (BETS)_2FeCl_4}$}}
\def\BETSGaCl{\mbox{${\rm (BETS)_2GaCl_4}$}}

\def\hsp#1{\hspace{#1ex}}

\def\Tc{{T_{\rm c}}}

\def\lsim{\stackrel{{\textstyle<}}{\raisebox{-.75ex}{$\sim$}}}
\def\gsim{\stackrel{{\textstyle>}}{\raisebox{-.75ex}{$\sim$}}}

\def\kF{k_{{\rm F}}}

\begin{document}
\draft 

\twocolumn[\hsize\textwidth\columnwidth\hsize\csname 
@twocolumnfalse\endcsname

\title{
Fulde-Ferrell-Larkin-Ovchinnikov State and Field-Induced \\
Superconductivity in an Organic Superconductor 
}

\author{Hiroshi Shimahara}

\def\runtitle{
FFLO State and Field-Induced Superconductivity 
in an Organic Superconductor 
}


\address{
Department of Quantum Matter Science, ADSM, Hiroshima University, 
Higashi-Hiroshima 739-8530, Japan
}

\date{Received 7 March 2002} 

\maketitle

\begin{abstract}
An experimental phase diagram of the field-induced superconductivity in 
the $\lambda$-\BETSFeCl \hsp{0.25} compound is theoretically reproduced 
by a combination of the Fulde-Ferrell-Larkin-Ovchinnikov (FFLO) state 
and the Jaccarino-Peter mechanism. 
Semi-quantitative agreement is obtained 
in the presence of a very weak subdominant triplet pairing interaction 
between electrons with antiparallel spins, 
in addition to a dominant singlet pairing interaction. 
An order-parameter mixing effect enhances the FFLO state 
and plays an essential role in the semiquantitative agreement. 
As a possible origin of the small triplet component, 
pairing interactions mediated by spin fluctuations are proposed. 
It is argued that, in this mechanism of the pairing interactions, 
equal spin pairing interactions are much weaker than the antiparallel 
spin pairing interactions in the strong magnetic field, 
which is consistent with the lack of observation of equal spin state. 
\end{abstract}



]

\narrowtext

Recently, 
a field-induced superconductivity (FISC) was observed 
at high magnetic fields 
in an organic superconductor $\lambda$-\BETSFeCl, 
(BETS = bis\-(ethylenedithio)\-tetra\-selena\-fulvalene)~\cite{Uji01}. 
The lower critical field $H_{\rm c}^{\rm low}(T)$ of the FISC 
is upwards convex~\cite{Bal01,Uji02} as a function of temperature $T$. 
When we consider the Jaccarino-Peter mechanism~\cite{Jac62} 
as the origin of the FISC~\cite{Bal01,Shi02c,Cep02}, 
the lower critical field corresponds to the upper critical field 
of a typical superconducting phase. 
Therefore, the upwards convex $H_{\rm c}^{\rm low}(T)$ 
is not a common characteristic of the upper critical field. 
Some authors argued~\cite{Bal01,Uji02,Shi02c} that this upturn 
behavior can be explained by the Fulde-Ferrell-Larkin-Ovchinnikov (FFLO) 
state~\cite{Ful64,Lar64,Bul74,Aoi74,Shi94a,Bur94}.

The FFLO state has been discussed in some other organic superconductors, 
because some of them satisfy the necessary conditions for the FFLO state to 
occur: (1) Those samples can be clean type II superconductors due to their 
narrow electron bands; (2) The orbital pair-breaking effect is strongly 
suppressed for parallel magnetic fields due to the low dimensionality. 
The present compound $\lambda$-\BETSFeCl \hsp{0.25} also exhibits 
these features.

The upturn due to the FFLO state in the cylindrically symmetric system 
was obtained by Bulaevskii~\cite{Bul74} 
and Aoi et al.~\cite{Aoi74}. 
The FFLO state in two-dimensional superconductors was reexamined and 
discussed in connection with exotic superconductors, 
such as organics and cuprates, by some 
authors~\cite{Shi94a,Bur94,Shi97a,Shi99,Shi02d,Mak96,Shi97b,Yan98,Shi00a}. 
The upturn is a common behavior of the FFLO state in quasi-two-dimensional 
systems~\cite{Shi97a,Shi99,Shi02d,Mak96,Shi97b,Yan98,Shi00a}. 
This behavior is due to a Fermi surface effect~\cite{Shi94a,Shi97a,Shi99} 
analogous to Fermi surface nesting 
for the spin-density and charge-density waves, 
which we call a nesting effect for the FFLO state.

On the other hand, an upturn of the upper critical field 
($\d^2 H_{\rm c}/\d T^2 > 0$) and 
a first-order transition below the upper critical field 
have been observed in $\kappa$-${\rm (BEDT}$-
${\rm TTF)_2Cu(NCS)_2}$~\cite{Nam99,Sin00,Sym01,Ish00}. 
The FFLO state has also been discussed in this compound. 
Manalo and Klein~\cite{Man00} have successfully confirmed the agreement 
between the experimental~\cite{Nam99,Sin00,Sym01} 
and theoretical results~\cite{Shi97b}.

A recent experiment on thermal conductivity by Tanatar et al.~\cite{Tan01} 
supports the existence of the FFLO state in $\lambda$-\BETSGaCl. 
They observed that the upper critical field is consistent with the 
Pauli paramagnetic limit in dirty samples, 
while it shows no saturation down to 0.3~K in clean samples. 
This experimental result supports the FFLO state also in the FISC 
in $\lambda$-\BETSFeCl, 
since these two compounds are similar, 
except $\lambda$-\BETSFeCl \hsp{0.25} has the magnetic anions.

In $\lambda$-\BETSFeCl, localized spins of $S = 5/2$ on the 
${\rm (FeCl_4)^{-1}}$ anions create an exchange field on the conducting 
layer of the two-dimensional network of ${\rm BETS}$ molecules. 
As many authors pointed out~\cite{Uji01,Bal01,Shi02c,Cep02}, 
the exchange field shifts the Zeeman energy effectively, and 
thus also the superconductivity area in the $B$-$T$ phase diagram. 
This compensation mechanism is called the Jaccarino-Peter mechanism, 
which was originally proposed for the FISC in ferromagnetic metals. 
The present compound is an antiferromagnetic insulator at zero field, 
but this difference does not affect the mechanism at high magnetic fields, 
where the localized magnetic moments are 
saturated~\cite{Bal01,Shi02c,Cep02}.

Balicas et al. have compared the experimental phase diagram of 
$\lambda$-\BETSFeCl \hsp{0.25} with the theoretical one with 
the FFLO state, taking into account the shift of the FISC area by 
the Jaccarino-Peter mechanism~\cite{Bal01}. 
The main features of the FISC have been reproduced qualitatively 
by a combination of the FFLO state and the Jaccarino-Peter mechanism.

However, from a quantitative viewpoint, 
the theoretical width of the magnetic field region of the FISC 
is much narrower than the experimental one. 
In the experimental phase diagram, the FISC area has a center at about 33~T 
and the lower limit at about 17~T~\cite{Uji02}, 
which means that the absolute value of the critical field is 
equal to about 16T in the absence of the exchange field. 
When the orbital pair breaking effect is ignored, 
the maximum transition temperature 
$T_{\rm c}^{\rm (m)} \approx 4.2~{\rm K}$ gives the Pauli paramagnetic limit 
$1.856{\rm [T/K]} \times T_{\rm c}^{\rm (m)} = 7.8~{\rm T}$ 
and the $s$-wave FFLO critical field 11~T for the cylindrical Fermi surface. 
Therefore, the experimental value 16~T is much larger than the estimates 
using the simple theory.

Therefore, we need an explanation for this difference. 
There are some mechanisms which enhance the FFLO critical field, 
such as the effects of the anisotropies of the Fermi surface and the gap 
function~\cite{Shi97a,Shi99,Shi02d,Mak96,Shi97b,Yan98}. 
In particular, by a nesting effect, the critical field can be enhanced 
up to a value several times as large as 
the Pauli paramagnetic limit~\cite{Shi97a,Shi99}. 
However, the increase due to this mechanism occurs 
at low temperatures~\cite{Shi02d}, 
and the tricritical point does not change. 
In the case of $\lambda$-${\rm (BETS)_2FeCl_4}$, 
it seems that the shape of the FISC area in the experimental phase 
diagram is not reproduced very well solely by these mechanisms.

Another mechanism is an order-parameter mixing effect. 
In the presence of a subdominant triplet pairing interaction 
between electrons with antiparallel spins, even if it is very weak, 
the upper critical field is largely enhanced. 
This effect was theoretically predicted in $s$-wave superconductors 
with a spherical symmetric Fermi surface~\cite{Mat94}, 
and also in $s$- and $d$-wave superconductors 
with quasi-two-dimensional Fermi surfaces~\cite{Shi00a}. 
In the latter study, this effect was discussed in connection with 
the organic superconductors. 
Subdominant triplet pairing components exist in 
the pairing interactions mediated 
by the spin fluctuations~\cite{And73,Shi00b}, 
and also in those mediated by phonons~\cite{Fou77}. 
In the latter, the triplet pairing component is enhanced due to 
the weakness of the screening effect 
in the low dimensions~\cite{Shi02b}.

In this paper, we reproduce a phase diagram that agrees with the 
experimental data semiquantitatively, 
by taking into account the order-parameter mixing effect in the FFLO 
state with the Jaccarino-Peter mechanism.

Before we analyze of the FISC, 
let us briefly discuss the exchange field created 
by the ${\rm (FeCl_4)}^{-1}$ anions. 
In our previous paper~\cite{Shi02c}, 
we proposed a mechanism which enhances the upper critical field 
in antiferromagnetic superconductors. 
In order to illustrate the mechanism, 
we used a generalized Kondo lattice model of 
the localized spins and mobile electrons 
with a Kondo coupling $J_{\rm K}$ between them 
and an antiferromagnetic exchange coupling $J$ 
between the localized spins. 
As a special case, the model describes an aspect of 
the $\lambda$-${\rm (BETS)_2FeCl_4}$ compound, 
for a half-filled electron band and large ratios of $J_{\rm K}/zJ$, 
where $z$ is the number of the nearest neighbor sites. 
In the application to the $\lambda$-${\rm (BETS)_2FeCl_4}$ compound, 
$J$ should be replaced with some indirect interaction via conducting 
layers~\cite{Cep02,Bro98}, which must be much smaller than $J_{\rm K}$. 
Therefore, it is expected that the ratio $J_{\rm K}/zJ$ is much smaller 
than 1. 
The RKKY exchange interaction and the reconstructed Fermi surface have 
been calculated by Brossard et al.~\cite{Bro98}.

In the calculation based on the generalized Kondo lattice 
model~\cite{Shi02c}, the boundary between the antiferromagnetic 
insulating phase and the paramagnetic metallic phase is 
$B_{\rm MI} = \mu_0 H_{\rm MI} = \mu_0 zJS/|\mu_{\rm e}|$, 
where $\mu_0$ is the magnetic permeability 
and $\mu_{\rm e}$ is the electron magnetic moment. 
On the other hand, the compensation effect shifts 
the center of the FISC area by 
\def\eqHcent{(1)}
$$
     B_{\rm cent} = \mu_0 H_{\rm cent} 
                  = \mu_0 \frac{J_{\rm K}S}{|\mu_{\rm e}|} . 
     \eqno\eqHcent
     $$
Therefore, from the experimental data~\cite{Uji01}, 
$B_{\rm MI} \approx 10.5{\rm T}$ and $B_{\rm cent} \approx 33{\rm T}$, 
we obtain the ratio 
\def\eqJKzJ{(2)}
$$
     \frac{J_{\rm K}}{zJ} = \frac{B_{\rm cent}}{B_{\rm MI}} 
          \approx 3.1 \gg 1 \hsp{0.5} , 
     \eqno\eqJKzJ
     $$
which is consistent with the above consideration.

From the values $B_{\rm cent} = 33~{\rm T}$, $B_{\rm MI} = 10.5~{\rm T}$ 
and $S = 5/2$, 
we obtain $J_{\rm K}/\kB \approx 8.87~{\rm K}$ and $zJ/\kB = 2.82~{\rm K}$. 
The mean field value of the antiferromagnetic transition temperature 
$T_{\rm AF}^{\rm MF} $ is equal to $zJS(S+1)/4 \approx 8.1~{\rm K}$, 
which agrees very well with the experimental value 8.5~K.

Now, let us examine the FISC. 
We examine the case in which 
a dominant singlet pairing interaction and a very weak subdominant 
triplet pairing interaction coexist 
between electrons with antiparallel spins. 
The pairing interaction and the gap function are expanded as 
\def\eqVandDeltaexpand{(3)}
$$
     \begin{array}{rcl}
     V(\vk,\vk') & = & \dps{ 
          - \sum_{\alpha} g_{\alpha} 
                     \gamma_{\alpha}(\vk) \gamma_{\alpha}(\vk') , 
          }\\
     \Delta (\vk) & = & \dps{ 
          \sum_{\alpha} \Delta_{\alpha} \gamma_{\alpha}(\vk) , 
          }
     \end{array}
     \eqno\eqVandDeltaexpand
     $$
with symmetry factors $\gamma_{\alpha}(\vk)$. 
In cylindrically symmetric systems, 
they are defined by, 
for example, 
$\gamma_{s}(\vk) \propto 1$, 
$\gamma_{p_x}(\vk) \propto {\hat k}_x$, 
$\gamma_{d_{x^2-y^2}}(\vk) \propto {{\hat k}_x}^2 - {{\hat k}_y}^2$, 
and so on. 
We normalize $\gamma_{\alpha}(\vk)$ by 
\def\eqgammanormlization{(4)}
$$
     \int_0^{2\pi} \frac{\d \varphi}{2\pi} 
          [\gamma_{\alpha}(\varphi)]^2 
          = 1 , 
     \eqno\eqgammanormlization
     $$
where $\gamma_{\alpha}(\varphi) = [\gamma_{\alpha}(\vk)]_{|\vk| = \kF}$ 
and $\varphi$ is the angle between $\vk$ and $k_x$-axis. 
We assume a cylindrically symmetric Fermi surface for simplicity, 
and $s$-wave and $p$-wave pairing interactions 
as the dominant and subdominant pairing interactions, 
respectively~\cite{Shi98b}. 
Later, we will discuss the effects of the anisotropies of 
the Fermi surface and the gap function.

In the weak coupling limit, the gap equations are written as 
\def\eqgapeqweakcoupling{(5)}
$$
     \Delta_{\alpha} \log \frac{T}{T_{{\rm c}\alpha}^{(0)}}
     = - \sum_{\beta} M_{\alpha \beta} \Delta_{\beta} , 
     \eqno\eqgapeqweakcoupling
     $$
where we define 
\def\eqMdef{(6)}
$$
\renewcommand{\arraystretch}{2.5}
     \begin{array}{rcl}
     M_{\alpha \beta} & \equiv & \dps{ 
          \int_0^{2\pi} 
          \frac{\d \varphi}{2\pi}
               \, \gamma_{\alpha}(\varphi) \gamma_{\beta}(\varphi) \, 
               \sinh^2 \frac{\beta \zeta}{2} \, \Phi(\varphi) , 
          } \\
     \Phi(\varphi) & \equiv & \dps{ 
          \int_0^{\infty} \hsp{-2} \d y \, 
               \frac{\tanh y}
                    {y \, (\cosh^2 y + \sinh^2 \frac{\beta \zeta}{2})} , 
          } \\
     T_{{\rm c}\alpha}^{(0)} & = & \dps{ 
          {2 \e^\gamma {\pi}^{-1} \omega_{\rm c}} 
          \e^{- 1/g_{\alpha} N(0)} , 
          } 
     \end{array}
     \eqno\eqMdef
     $$
with 
$\zeta = h \, [\qbar \cos (\varphi - \varphi_{\vq}) + 1] $, 
$\qbar = \vF q/2h$, 
$h = |\mu_e| (|H| - H_{\rm cent})$, 
$N(0)$ the density of states at the Fermi energy, 
and $\varphi_{\vq}$ the angle between the center-of-mass momentum $\vq$ 
of the FFLO state and the $k_x$-axis.

For the symmetry, we can assume the direction of $\vq$ 
along the $k_x$-axis without loosing generality. 
Then, the $p_y$-wave component $\Delta_{p_y} {\hat k}_y$ 
is not mixed with the $s$-wave component, 
while the $p_x$-wave component $\Delta_{p_x} {\hat k}_x$ 
is mixed with it. 
Therefore, the transition temperature for a given $\vq$ is calculated by 
$\Tc(h,\vq) = T_{{\rm c}s}^{(0)} \e^{-\lambda(h,\vq)}$ 
with the smallest eigenvalue 
$\lambda(h,\vq)$ of the $2 \times 2$ matrix 
\def\eqspmixingMatrix{(7)}
$$
     {\left (
     \begin{array}{cc}
     M_{ss} & M_{sp} \\
     M_{ps} & M_{pp} + G_{p} 
     \end{array}
     \right )} , 
     \eqno\eqspmixingMatrix
     $$
where we define 
\def\eqGpdef{(8)}
$$
     G_{p} \equiv \log \frac{T_{{\rm c}s}^{\rm (0)}}{T_{{\rm c}p}^{(0)}} 
     = \frac{1}{g_{p}N(0)} - \frac{1}{g_{s}N(0)} . 
     \eqno\eqGpdef
     $$
Here, $T_{{\rm c}s}^{(0)}$ and $T_{{\rm c}p}^{(0)}$ are the pure $s$- and 
$p$-wave transition temperatures. 
In the presence of the exchange field, 
$T_{{\rm c}s}^{(0)}$ is equal to the maximum transition temperature 
$T_{\rm c}^{\rm (m)}$ of the FISC. 
The final result of the transition temperature is obtained by maximizing 
$T_{\rm c}(h,\vq)$ with respect to $\vq$.

\begin{figure}[htb]
\begin{center}
\begin{tabular}{c}
\leavevmode \epsfxsize=7cm  
\epsfbox{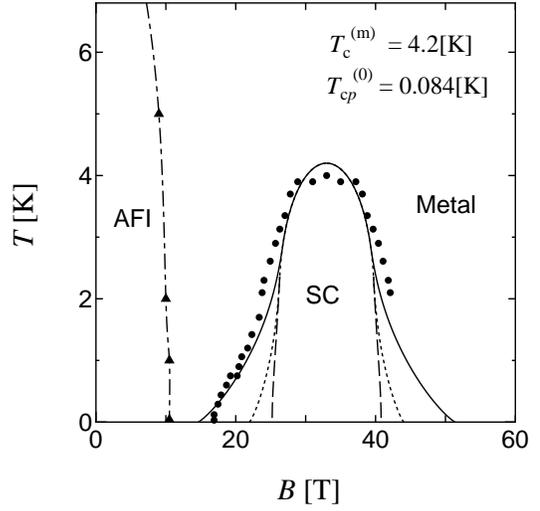}
\\[24pt]
\leavevmode \epsfxsize=7cm  
\epsfbox{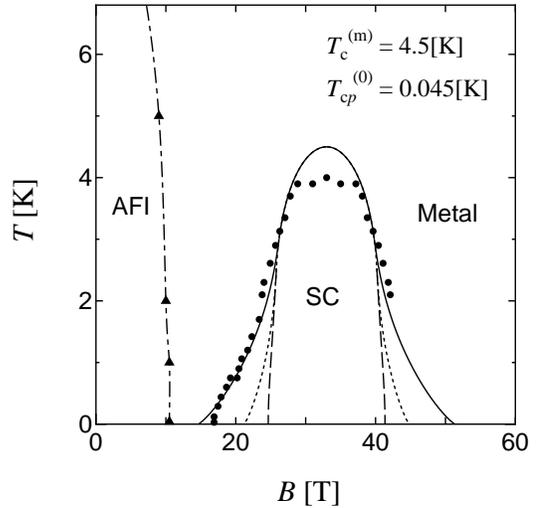}
\end{tabular}
\end{center}
\caption{
Comparison between the theoretical and experimental phase diagrams 
in $\lambda$-${\rm (BETS)_2FeCl_4}$. 
We set the maximum value of the superconducting transition temperature 
(a) $T_{\rm c}^{\rm (m)} = 4.2~{\rm K}$ 
and $T_{{\rm c}p}^{(0)} = 0.084~{\rm K}$, 
and 
(b) $T_{\rm c}^{\rm (m)} = 4.5~{\rm K}$ 
and $T_{{\rm c}p}^{(0)} = 0.045~{\rm K}$. 
The strength of the exchange field is assumed 
so that the center of the peak $B_{\rm cent}$ 
is equal to 33~T. 
The solid, dotted, and dashed curves show boundaries of the FISC, 
in the presence of the triplet pairing interaction, 
in the absence of it, 
and when the FFLO state is excluded, respectively. 
The solid circles and the solid triangles show 
the experimental data of the superconducting 
and antiferromagnetic transition points, respectively, 
by Uji et al. in ref.~[3]. 
The dot-dashed curve is a guide for the eyes. 
} 
\label{fig:phased}
\end{figure}

The results are depicted in Fig.~\ref{fig:phased}. 
The FFLO critical fields are largely enhanced even for a very small value 
of the coupling constant of the coexisting triplet pairing interaction. 
Figure~\ref{fig:phased}(a) shows the result for 
$T_{\rm c}^{\rm (m)}= 4.2~{\rm K}$ 
and $T_{{\rm c}p}^{(0)} = 0.084~{\rm K}$, 
which corresponds to 
$T_{{\rm c}p}^{(0)}/T_{\rm c}^{\rm (m)} = 1/50$ and 
$G_p \approx 3.912$. 
It is found that 
the agreement between theoretical and experimental results is significantly 
improved by taking into account the weak triplet pairing interaction 
(solid curve), in comparison to the result in the absence of the 
triplet pairing interactions (dotted curve). 
Taking into account the ambiguity in determination of 
the experimental values of 
the transition temperatures~\cite{Uji01,Bal01,Uji02}, 
it can be concluded that 
the overall behavior of the transition curve is semiquantitatively 
reproduced by the FFLO state.

Figure~\ref{fig:phased}(b) is the result for 
$T_{\rm c}^{\rm (m)}= 4.5~{\rm K}$ 
and $T_{{\rm c}p}^{(0)} = 0.045~{\rm K}$, 
which corresponds to 
$T_{{\rm c}p}^{(0)}/T_{\rm c}^{\rm (m)} = 1/100$ and 
$G_p \approx 4.605$. 
Except near the top of the peak, 
the theoretical curve agrees very well with the experimental data.

In the above calculations, the presence of the weak triplet pairing 
interactions between electrons with antiparallel spins is assumed. 
When the system is isotropic in the spin space, 
triplet pairing interactions between electrons with parallel spins 
are of the same magnitude at zero field. 
In this case, one might expect transitions to an equal spin pairing 
state at magnetic fields where the FISC of antiparallel spin pairing 
is suppressed. 
For example, in the cases shown in Figs.~\ref{fig:phased}(a) 
and \ref{fig:phased}(b), 
the transition temperatures of the equal spin state are equal 
to 84~mK and 45~mK, respectively. 
However, such transitions have not been observed in the experiments.

\vspace{\baselineskip}

\begin{figure}[htb]
\begin{center}
\begin{tabular}{lc}
(a) & \hsp{2} 
\begin{tabular}{c}
\leavevmode \epsfxsize=4cm  
\epsfbox{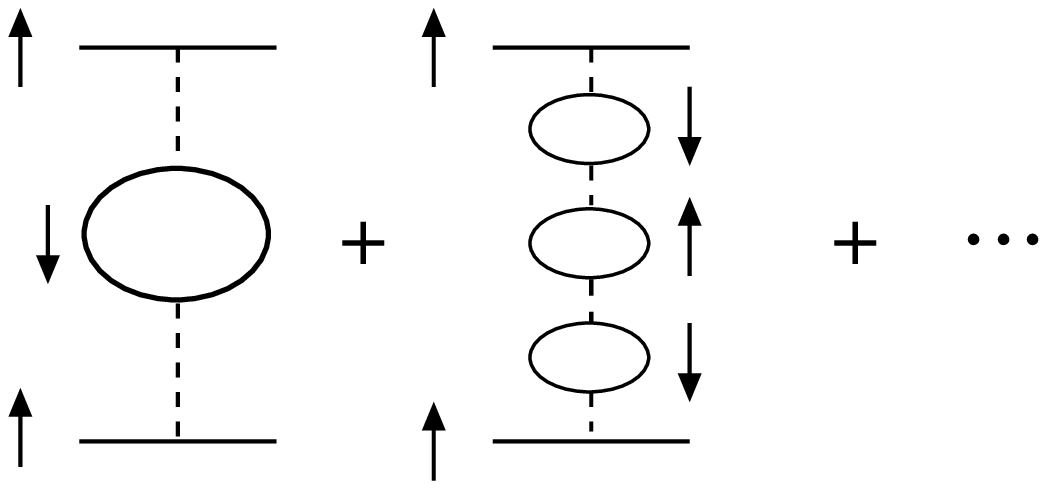} 
\end{tabular}
\\[36pt]
(b) & \hsp{2} 
\begin{tabular}{c}
\leavevmode \epsfxsize=4cm  
\epsfbox{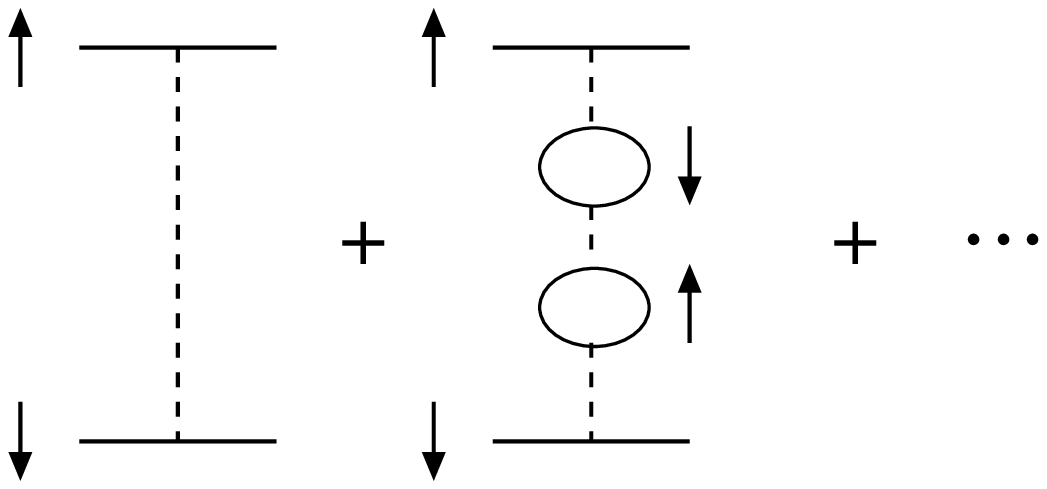} 
\end{tabular}
\\[36pt]
(c) & \hsp{2} 
\begin{tabular}{c}
\leavevmode \epsfxsize=5cm  
\epsfbox{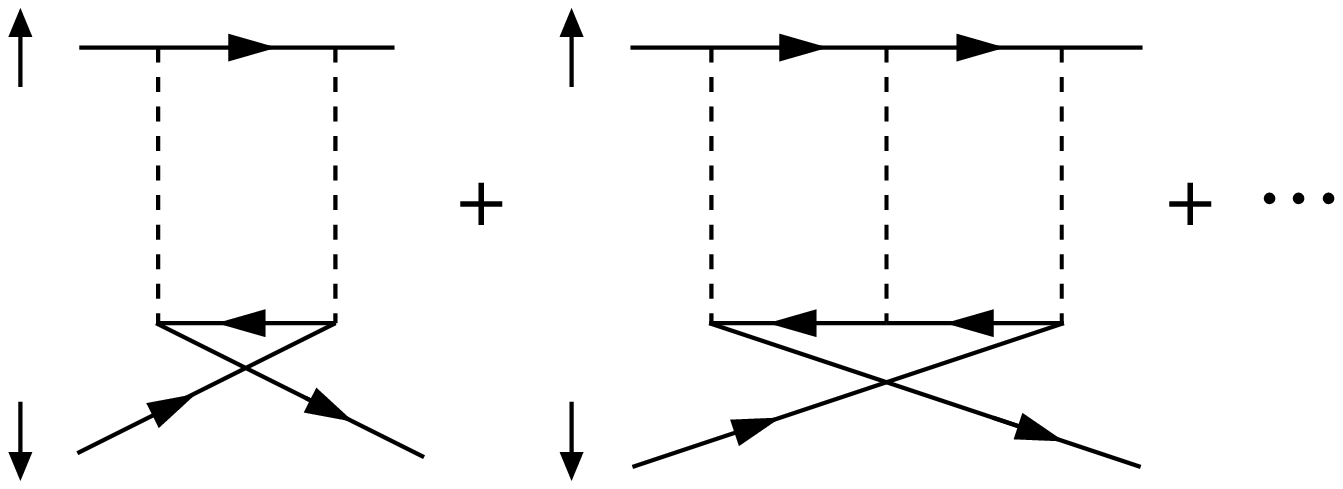} 
\end{tabular}
\end{tabular}
\end{center}
\caption{
Series of diagrams of fluctuations 
which contribute to the pairing interactions. 
The short dashed line denotes the on-site Coulomb interaction $U$. 
} 
\label{fig:diagrams}
\end{figure}

This discrepancy can be explained 
if the origin of the triplet pairing interactions is the exchange of 
the spin fluctuations enhanced by on-site Coulomb 
repulsion~\cite{And73,Ber66}. 
For example, in a random phase approximation (RPA), 
the equal spin pairing interactions originate from 
the fluctuations as shown in the diagrams in 
Fig.~\ref{fig:diagrams}(a)~\cite{And73}. 
Therefore, they correspond to the fluctuations of $S_z$ components. 
On the other hand, 
the antiparallel spin pairing interactions originate 
from the fluctuations as shown in the diagrams in 
Figs.~\ref{fig:diagrams}(b) and \ref{fig:diagrams}(c)~\cite{And73,Ber66} 
in the RPA. 
They correspond to the fluctuations of $S_z$ components and 
$S_{\pm} \equiv S_x \pm \i S_y $, respectively. 
In the magnetic field in the $z$-direction, 
since the spin moment $\langle S_z \rangle$ is finite, 
the spin fluctuations of $z$ components are much smaller than 
those of $x$ and $y$ components. 
Therefore, the equal spin pairing interactions are much weaker 
than the antiparallel spin pairing interactions in the magnetic field.

\vspace{\baselineskip}

\vspace{\baselineskip}

Another possible explanation is the orbital pair-breaking effect. 
The transition temperature of the equal spin pairing state is 
only 45~mK in the case shown in Fig.~\ref{fig:phased}(b). 
If we take into acount the orbital pair-breaking effect, 
the equal spin pairing state with such a small transition temperature 
may be suppressed completely at higher magnetic fields 
where the system is metallic.

In this paper, we have assumed the $s$-wave pairing interaction 
as the dominant singlet interaction, and the cylindrical Fermi surface 
for simplicity~\cite{Shi98b}. 
For the $d$-wave pairing, 
the curve of the critical field exhibits a kink 
at a low temperature~\cite{Mak96,Shi97b,Yan98}. 
However, the kink disappears when the Fermi surface is anisotropic 
to some extent~\cite{Shi02d}. 
Furthermore, the anisotropy of the Fermi surface enhances 
the critical field~\cite{Shi97a,Shi99} for a nesting effect. 
However, since the enhancement occurs especially at low temperatures 
$T \lsim 0.2 \Tc$ ~\cite{Shi02d}, 
the mixing effect remains necessary for the semiquantitative reproduction 
of the experimental phase diagram in an entire temperature region 
including $T \gsim 0.2 \Tc$.

In conclusion, the phase diagram of $\lambda$-${\rm (BETS)_2FeCl_4}$ 
including the curve of the lower limit of the FISC which is 
upwards convex ($\d^2 H_c(T)/\d T^2 < 0$) 
can be explained by the combination of 
the FFLO state and the Jaccarino-Peter mechanism, together with the 
order-parameter mixing effect. 
For the semiquantitative reproduction of the phase diagram in the 
present theory, 
the existence of the very weak subdominant triplet pairing interaction 
is essential. 
The origin of it can be attributed to spin fluctuations 
in the magnetic field. 
The phase diagrams of $\lambda$-${\rm (BETS)_2Ga_{1-x}Fe_xCl_4}$ are 
also expected to be understood in the present mechanism. 
The experimental evidence of the existence or inexistence of the FFLO 
state can be obtained by direct observations of the structure of 
the gap function~\cite{Shi98a,Kle00}.

The author would like to thank Dr.~Uji for useful discussions and 
the experimental data, and Dr.~C${\rm {\acute e}}$pas 
for useful discussions.




\begin{thebibliography}{99}
\bibitem{Shi98b} 
  It was argued that the FFLO states with one dimensional structures are 
  not stable against the phase fluctuations in isotropic systems, 
  but they are stabilized by anisotropy of the Fermi surface, 
  [H.~Shimahara: Physica B {\bf 259-261} (1999) 492; 
  J. Phys. Soc. Jpn. {\bf 67} (1998) 1872; Y.~Ohashi: preprint]. 
  However, since there is anisotropy of some extent in the real materials, 
  this fluctuation effect can be neglected practically, 
  and as far as it is neglected, the isotropic model can be used 
  for the estimation of $T_{\rm c}$ as an approximation.
\end{thebibliography}
\end{document}